\documentclass[apjl]{emulateapj}

\usepackage{amssymb}
\usepackage{natbib}
\usepackage{graphicx}
\usepackage{color}
\usepackage{subfigure}

\def\aap{A\&A }
\def\mnras{MNRAS }
\def\apj{ApJ }
\def\pasa{Publ. Astron. Soc. Aust.}
\def\rmxaa{Rev. Mexicana Astron. Astrofis.}

\shorttitle{The Combined Impact of Ionizing Radiation and Momentum Winds on a Self-Gravitating Core}
\shortauthors{J. Ngoumou, D. Hubber,  J. Dale, A. Burkert}

\begin{document}

\title{FIRST INVESTIGATION OF THE COMBINED IMPACT OF IONIZING RADIATION AND MOMENTUM WINDS FROM A MASSIVE STAR ON A SELF-GRAVITATING CORE }

\author{Judith Ngoumou\altaffilmark{1$^*$}, David Hubber\altaffilmark{1,2}, James E. Dale\altaffilmark{1,2}, and Andreas Burkert \altaffilmark{1}}
\affil{$^{1}$Universit\"ats-Sternwarte M\"unchen, Ludwig-Maximilians-Universit\"at, Scheinerstr.1, 81679 M\"unchen, Germany\\}
\affil{$^{2}$Excellence Cluster ÔUniverseÕ, Boltzmannstr. 2, 85748 Garching, Germany}
\email{$^*$ngoumou@usm.uni-muenchen.de}

\begin{abstract}

Massive stars shape the surrounding interstellar matter (ISM) by emitting ionizing photons and ejecting material through stellar winds. To study the impact of the momentum from the wind of a massive star on the surrounding neutral or ionized material, we  implemented a new HEALPix-based momentum-conserving wind scheme in the smoothed particle hydrodynamics (SPH) code SEREN. A qualitative study of the impact of the feedback from an O7.5-like star on a self-gravitating sphere shows that on its own, the transfer of momentum from a wind onto cold surrounding gas has both a compressing and dispersing effect. It mostly affects gas at low and intermediate densities.
When combined with a stellar source's ionizing ultraviolet (UV) radiation,  we find the momentum-driven wind to have  little direct effect on the gas. We conclude that, during a massive star's main-sequence, the UV ionizing radiation is the main feedback mechanism shaping and compressing the cold gas. Overall, the wind's effects on the dense gas dynamics and on the triggering of star formation are very modest. The structures formed in the ionization-only simulation and in the combined feedback simulation are remarkably similar. However, in the combined feedback case, different SPH particles end up being compressed. 
This indicates that the microphysics of gas mixing differ between the two feedback simulations and that the winds can contribute to the localized redistribution and reshuffling of gas.

\end{abstract}

%methods: numerical --- ISM: bubbles --- stars: winds, outflows --- ISM: clouds --- (ISM:) HII regions --- stars: massive}

\section{Introduction}
 \label{sec:intro}
 
During their lifetime, stars with masses greater than $8\,M_\sun$ influence their surroundings by injecting energy, mass, and momentum through feedback mechanisms such as ionizing radiation, stellar winds, or radiation pressure. 
Observations and numerical simulations have shown that the above feedback mechanisms can produce a variety of  structures including superbubbles at large scales \citep{2004ApJ...613..302O, 2011ApJ...731...13N},  cavities \citep{2012ASPC..453...25F},  shells \citep{2010A&A...523A...6D, 2012MNRAS.427..625W}, pillars and filaments \citep{2010ApJ...723..971G, 2011A&A...525A..92P, 2012MNRAS.427..625W}, and bow-shocks seen around moving stars \citep{ 2012AJ....143...71K, 2012MNRAS.424.3037G,2012A&A...541A...1M,2013MNRAS.436..859M, 2013ApJ...769..139N}. 

A sizable effect of the feedback mechanisms at work during the lifetime of the massive stars (before the final supernova explosion) is to fill voids and leak out of highly structured clouds \citep{2011MNRAS.414..321D, 2014MNRAS.442..694D}. 
\citet{2011ApJ...735...66M} argue that radiation pressure in massive clusters is a viable mechanism to expel gas and launch super-galactic winds. 

Feedback processes can also affect star formation locally by either triggering the formation of new stars \citep{2010ApJ...723..971G, 2012MNRAS.427..625W, 2012A&A...540A..81O} or dispersing clouds and thereby delaying or even hindering star formation \citep[][]{2011MNRAS.417..950H,2012MNRAS.427..625W, 2013MNRAS.436.3430D}.  

Theoretical studies have examined the impact of massive star feedback on 
uniform density surrounding media and derived analytical descriptions for the evolution and structures of spherical H\,{\sc ii} regions and stellar wind bubbles. The pioneering work of \cite{1939ApJ....89..526S} laid the foundation for understanding the formation of H\,{\sc ii} regions, paving the way for the derivation of the time evolution of an ionization front \citep{1978ppim.book.....S}.  

The evolution of stellar wind bubbles has also been studied in great detail analytically \citep[][]{1975ApJ...200L.107C, 1977ApJ...218..377W, 1988RvMP...60....1O, 2001PASP..113..677C}. In the classical picture, the  wind bubble expansion into a uniform medium during the main-sequence stage can be divided into three stages \citep{1999isw..book.....L}. The first two phases, the free-expansion phase and the fully adiabatic phase are of very short duration ( $\sim 10^2\,\rm{yr}$  and $\sim 10^3\,\rm{yr}$ respectively; \citealt{1999isw..book.....L}). The third phase, the snowplow phase, is the longest  ( $\ge 10^6 \,\rm{yr}$) and lasts for most of the star's main-sequence life and is therefore more likely to be observed. This phase describes the evolution of a cold ($T \leq 10^4 \,\rm{K}$) shell of swept-up interstellar gas, encompassing the shocked wind material. Depending on whether or not the shocked wind region has cooled, one can distinguish between the ''energy conserving''  snowplow regime with a shell expansion law of $R_{_{\rm SHELL}} \propto t^{3/5}$ \citep[see e.g.][]{1975ApJ...200L.107C}, and the ''momentum-conserving'' snowplow expansion  with  $R_{_{\rm SHELL}} \propto t^{1/2}$ \citep[see e.g.][]{1975ApJ...198..575S}.

The question of whether or not the hot interior is able to cool has not been fully answered yet.  \citet{1977ApJ...218..377W} showed, assuming that the shocked wind region is delimited by a collisionless shock at the interior and by an insulating contact discontinuity from the outside, that the hot wind material would mainly cool by adiabatic expansion. However this could only happen on timescales longer than the main-sequence life of the star. The inclusion of  thermal conduction effects at the contact discontinuity do not lead to a drastically different expansion law. However, discarding the assumption of a contact discontinuity and assuming effective mass loading and mixing between the shocked wind and the ambient material could lead to effective cooling in the bubble \citep{2001PASP..113..677C}. 

Two-dimensional (2D) numerical simulations have modeled the expansion of an energy-driven bubble and the evolution of circumstellar material  from the main-sequence to the Wolf-Rayet phase \citep{1995ApJ...455..145G, 1996A&A...305..229G,1996A&A...316..133G}. They show that hydrodynamical instabilities can develop during the evolution of wind-blown bubbles.  
\cite{2003ApJ...594..888F,2006ApJ...638..262F} included the effect of ionizing radiation and presented the picture of a wind bubble contained inside an H~{\sc ii} region. 
\cite{2009ApJ...693.1696H} modified the classical \citeauthor{1975ApJ...200L.107C} and   \citeauthor{1977ApJ...218..377W}
theory to take into account the gas leakage in their model of the Carina Nebula.  They argue that  the shocked stellar winds are not important for the dynamical evolution of the bubble.

From an observational point of view wind-blown bubbles have rarely been observed around main-sequence massive stars \citep{2007dmsf.book..183A}. The  diffuse X-ray emission predicted by models is in disagreement with the rare observational detection of soft X-ray bubbles around main-sequence massive stars \citep{2006ESASP.604..363C}, hinting at the possibility of cooler main-sequence bubbles \citep{2000RMxAC...9..273M}.

In this paper we investigate the effects of the momentum-driven winds and their interplay with ionizing stellar radiation. We present the implementation of a numerical method for three-dimensional (3D) hydrodynamical simulations, which allows for the injection of momentum imparted by a constant, isotropic stellar wind using the HEALPix tessellation scheme \citep{2005ApJ...622..759G}.  Our method can be used in conjunction with the ionizing radiation scheme from \cite{2009A&A...497..649B}. Section~\ref{sec:num} briefly describes the implementation of the method in the smoothed particle hydrodynamics (SPH) code SEREN \citep{2011A&A...529A..27H}. We apply the new scheme to model the momentum-conserving phase of wind front expansion in a cold uniform. In Section~\ref{sec:ids} we examine the combined effect of the momentum transfer and the ionizing radiation as implemented by  \cite{2009A&A...497..649B}. In section~\ref{sec:bes} we examine the effect of wind and ionization feedback on a self-gravitating core. We present our conclusions in Section~\ref{sec:conc}.

%%%%%%%%%%%%%%%%%%%%%%%%%%%%%%%%%%%%%%%%%%%%%%%%%%%%%%%%%%%%
\section{Numerical Scheme}
 \label{sec:num}

\subsection{Implementation Method} \label{subsec: method}

 \begin{figure}[h]
 \centering
\includegraphics[angle= 0, width = 0.85\linewidth]{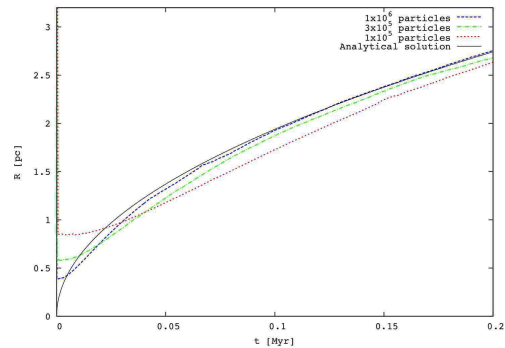}
\caption{ Evolution of the wind shell position with time. Comparison between simulations with  $1\times10^5$ (red dotted), $3\times10^5$ (green dot-dashed) particles and $1\times10^6$ particles (blue dashed). The black line follows the analytical prediction given by Equation~\ref{mix_R2}}.
\label{resolution1}      
\end{figure}

 \begin{figure*}[ht]
\includegraphics[width = \linewidth]{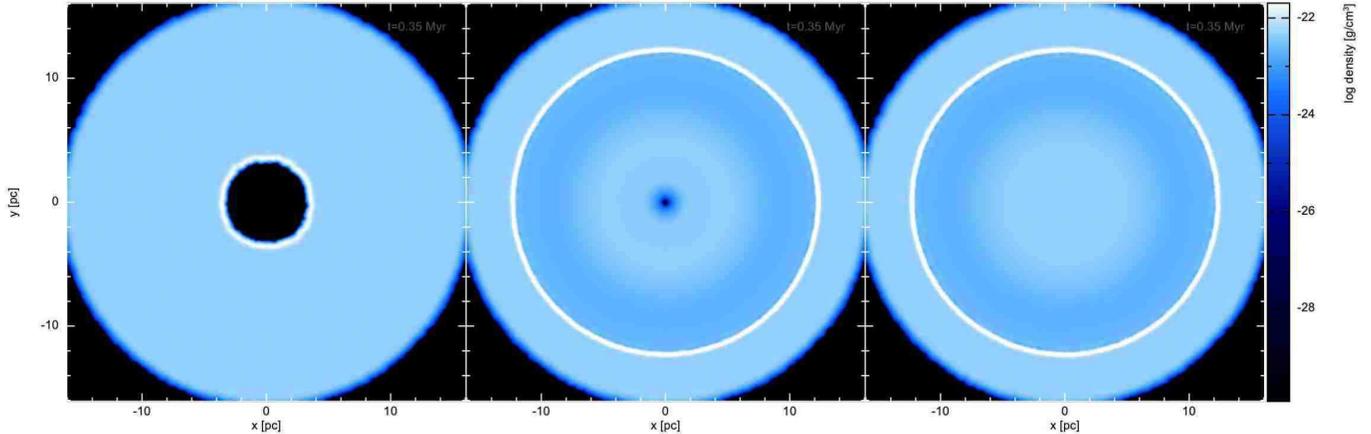}
\caption{Density slice through a column density plot showing the shell expansion in a uniform density medium at the same time $t=0.35\,{\rm Myr}$ for three feedback mechanisms.
	Left:  momentum transfer only. 
	Middle:  momentum transfer and ionizing radiation.
	Right:  ionizing feedback only}
\label{D0winion}      
\end{figure*} 

Assume that a star located at position ${\bf r}_{_{\rm STAR}}$ is emitting an isotropic mechanical wind at a mass-loss rate $\dot{M}$ and a wind speed $v_{_{\rm WIND}}$.  The rate of total (scalar) linear momentum carried by the wind is 

\begin{equation}
\dot{p}_{_{\rm WIND}} = \dot{M}\,v_{_{\rm WIND}}.\\
\end{equation}
We use the HEALPix algorithm \citep[][]{2005ApJ...622..759G} to split the spherical surface surrounding the source into discrete elements covering approximately equal areas, which allows us to discretize the wind emitted by the star.  In HEALPix, the first level of rays ($l = 0$) contains $12$ discrete rays.  For increased resolution each subsequent level is achieved by splitting the rays into 4 child rays. The number of rays on each level $l$ is given by ${\cal N}_{_{\rm RAYS}} = 12 \times 4^l$.  Per design, the HEALPix algorithm allows for a maximum level of refinement $l_{max}\leq 12$. In our present study, we use $l_{max}=7$. 
At a given time, a ray on level $l$ carries a momentum package  given by the momentum rate

\begin{equation}
\dot{p}_l = \frac{\dot{M}\,v_{_{\rm WIND}}}{12 \times 4^l}.
\end{equation}

For each feedback source a linked list of particles sorted by increasing distance from the feedback source is constructed along each ray on the first level.  As we walk the HEALPix rays, we find the first SPH particle on the ray with a smoothing length $h_{_{\rm FIRST}}$ at a distance  $d_{_{\rm FIRST}}  = |{\bf r}_{_{\rm FIRST}} - {\bf r}_{_{\rm STAR}}| $ from the star.

 We then check if the ray resolution is acceptable, i.e. if the separation between neighboring rays is less than 
  the smoothing length $h_{_{\rm FIRST}}$. This is given by the splitting criterion described in \citet[][]{2009A&A...497..649B} and controlled by a dimensionless parameter $f_2$ which sets the angular resolution of the rays. If  $d_{_{\rm FIRST}} \Delta\theta_l > f_2 h_{_{\rm FIRST}}$, $ \Delta\theta_l$ being the angle between neighboring rays at level $l$, the ray is split into 4 new child rays. This procedure is repeated for the child rays until the required resolution is reached.
We use $f_2 = 0.5$ for all the simulations reported in this paper.

We then walk the list up the ray until we find all SPH particles contained  
between $|{\bf r}_{_{\rm FIRST}} - {\bf r}_{_{\rm STAR}}|$ and  $|{\bf r}_{_{\rm FIRST}} - {\bf r}_{_{\rm STAR}}| + {\cal R}\,h_{_{\rm FIRST}}$ of the source, where ${\cal R}$ is the compact support of the SPH kernel function (e.g. ${\cal R} = 2$ for M4-kernel).  The momentum is distributed only among the first  particle in each ray and its immediate neighbors (within ${\cal R}\,h_{_{\rm FIRST}}$).
We calculate the acceleration of these particles by distributing the momentum flux belonging to that ray among them.  In order to account for the geometric dilution of the wind as the radius increases, we weight the accelerations given to each particle by $r^{-2}$.  Therefore the rate of change of linear momentum due to the wind for particle $i$ is given by 

\begin{equation}
\dot{p}_i = \frac{\dot{M}\,v_{_{\rm WIND}}}{12 \times 4^l} \,
\frac{m_i \vert {\bf r}_i - {\bf r}_{_{\rm STAR}} \vert ^{-2}}
{\sum \limits_{j=1}^{N} m_j\,\vert{\bf r}_j - {\bf r}_{_{\rm STAR}}\vert^{-2}},\\
\label{mom_rate_i}
\end{equation}
 where the summation is over all particles between $|{\bf r}_{_{\rm FIRST}} - {\bf r}_{_{\rm STAR}}|$ and $|{\bf r}_{_{\rm FIRST}} - {\bf r}_{_{\rm STAR}}| + {\cal R}\,h_{_{\rm FIRST}}$ in that ray. The sum is used to normalize the total wind momentum in the selected ray. 

From Newton's second law, we get

\begin{equation}
\frac{d}{dt}(p_i) = \frac{dm_i}{dt}v_i + \frac{dv_i}{dt}m_i.\\
\end{equation}
Therefore, if we assume that the mass of the wind is negligible (i.e. $\frac{dm_i}{dt} \times \Delta t \ll m_i$), then the first term on the right is negligible and the rate of change of momentum is given by 

\begin{equation}
{\bf a}_i = \frac{\dot{p}_i}{m_i} \, 
\frac{{\bf r}_i - {\bf r}_{_{\rm STAR}}}
{|{\bf r}_i - {\bf r}_{_{\rm STAR}}|}.
\end{equation}
As explained in \citet{2007ApJ...671..518K}, the ray ensemble is rotated about three random angles to avoid numerical artifacts that might appear at the border of  the rays due to the angular discretization.

 \subsection{Expansion in a Cold Uniform Medium} 
\label{subsec:udscold}

The expansion of a wind bubble shell in a uniform density medium with negligible pressure during the momentum-conserving snowplow phase can easily be derived \citep[see e.g.][]{1975ApJ...198..575S, 1988RvMP...60....1O, 1999isw..book.....L}. 
 The equation of momentum conservation is given by

\begin{equation}
\frac{d}{dt} \left(M_{_{\rm SHELL}} v_{_{\rm SHELL}}\right) = 4\pi R_{_{\rm SHELL}}^2 p_{_{\rm WIND}}.\\
\label{mom1}
\end{equation}
where $p_{_{\rm WIND}} =  \rho_{_{\rm WIND}} v_{_{\rm WIND}}^2$ is the ram pressure of the wind and $\rho_{_{\rm WIND}}=\dot{M}/(4\pi R_{_{\rm SHELL}}^2 v_{_{\rm WIND}})$ is the wind density.  $\dot{M}$ and $v_{_{\rm WIND}}$ are the stellar mass loss rate and the wind terminal velocity respectively.  The mass of the swept up shell is $M_{_{\rm SHELL}}= (4/3)\pi R_{_{\rm SHELL}}^3 \rho_0$ with $\rho_0$ being the initial undisturbed density of the gas.
We obtain

\begin{equation}
\frac{\pi}{3}\rho_0 \frac{d^2(R_{_{\rm SHELL}}^4)}{dt^2}  =  \dot{M} v_{_{\rm WIND}}. 
\label{mix_mom}
\end{equation}
Assuming  a power-law form for the solution $R_{_{\rm SHELL}} \propto t^{\gamma}$, 
The solution to Equation~\ref{mix_mom} can be calculated as:

\begin{equation}
R_{_{\rm SHELL}}(t) = 0.83\dot{M}^{1/4} v_{_{\rm WIND}}^{1/4} \rho_0^{-1/4} t^{1/2} 
\label{mix_R2}
\end{equation} 

  \begin{figure}[!h]
 \centering
\includegraphics[angle=0, width = 0.85\linewidth]{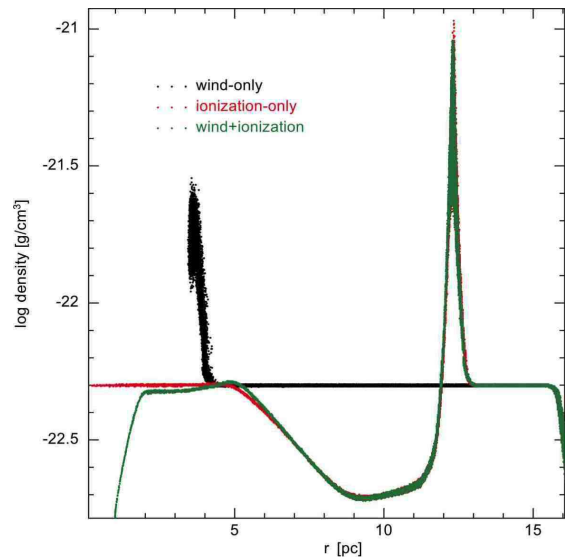}
\caption{Density over radial distance $r$  from the feedback source for the wind-only simulation (black dots), the ionization-only simulation (red dots) and the combined feedback case (green dots) at the same time $t=0.35\,{\rm Myr}$.}
\label{uds_rho_r}      
\end{figure}
%  \begin{figure*}[!p]
% \hspace{-2.0em}
%\includegraphics[width =1. \linewidth]{BES5e5_c018_env_w_O7_3x3.eps}
%\caption{Time  evolution of the impact of the momentum transfer on a cold core. The color bar represents the integrated density along the z-axis in $\rm{g/cm^{2}}$ }
%\label{windcore}      
%\end{figure*} 

We performed a set of simulations with  $1\times10^5$ (particle mass $m_{_{\rm PART}} = 4.13\times10^{-2}\,{\rm M}_{\odot}$), $3\times10^{5}$ ($m_{_{\rm PART}} = 1.4\times10^{-2}\,{\rm M}_{\odot}$),  $1\times10^6$ particles ($m_{_{\rm PART}} = 4.13\times10^{-3}\,{\rm M}_{\odot}$).
 Our  cloud is modeled as a spherical uniform density cloud of density ${n}_{\rm c} = 30\,{\rm cm}^{-3}$ and temperature $T =10$~K. The wind source is located at the center. We used fixed values for the wind mass loss $\dot{{ M}} = 10^{-6}\,{\rm M}_{\odot}\, {\rm yr}^{-1}$ and the wind velocity ${v}_{_{\rm WIND}} = 2000\,{\rm km}\, {\rm s}^{-1}$. 
 The transfer of momentum leads to the formation of a shock front which expands and sweeps over the material surrounding the source. We used the mean of the positions of the $100$ densest particles to identify the position $R_{_{\rm SHELL}}$ of the shock front in our simulations.
 
In  Figure~\ref{resolution1} we compare the theoretical expansion law (Equation~\ref{mix_R2}) with the shock front evolution obtained in our simulations. The inaccuracies seen at the beginning are related to the initial smoothing length $h$ since we smooth the momentum over $2h$ as described in Section~\ref{subsec: method}.
With increasing resolution the shock front expansion converges towards the analytical solution. Runs with more than  $3\times10^5$  particles are in good agreement with the analytical expectation.
Unlike the ionization scheme from \citet{2009A&A...497..649B}, the momentum wind implementation does not require additional temperature smoothing and thus is a robust representation of the physics involved.

%%%%%%%%%%%%%%%%%%%%%%%%%%%%%%%%%%%%%%%%%%%%%%%%%%%%%%%%%%%%

\section{Impact of the Momentum Transfer on an Ionized Uniform Cloud}
 \label{sec:ids}

In order to assess the impact of the wind on the surroundings of the star,
the momentum transfer scheme was applied to a uniform density cloud, ${n}_{\rm c} \approx 30\,{\rm cm}^{-3}$. We used fixed values for the wind mass loss $\dot{{ M}} = 10^{-6}\,{\rm M}_{\odot}\, {\rm yr}^{-1}$,  the wind velocity $v_{_{\rm WIND}}  = 2000\,{\rm km}\, {\rm s}^{-1}$ and the ionizing photon rate  ${ N}_{{\rm Lyc}} = 10^{49}\,{\rm s}^{-1}$. These are values close to those for an O7.5 star as listed by \citet{2006MNRAS.367..763S}.
The effects of ionizing radiation are included using the HEALPix based ionizing radiation scheme developed in \cite{2009A&A...497..649B}.
The results are compared for these three cases of stellar feedback in Figure~\ref{D0winion}.

Figure~\ref{uds_rho_r} shows the radial density profile at  $t=0.35\,\rm{Myr}$ for the wind-only simulation (black), the ionization-only case (red) and the combined feedback case (green). 
In a cold uniform medium, the momentum transfer from a single stellar wind creates an expanding shock front. Its impact on warm ionized material is however significantly reduced. The pressure in the $10^4$~K environment is high enough to decelerate the front until it reaches the sound speed in the ionized gas and a quasi equilibrium is attained as the ram pressure equals the thermal pressure in the ionized gas. 
A nearly stable configuration is achieved with just the innermost $\sim 1$~pc affected by the momentum wind.

\begin{figure}[ht]
  \hspace{-3mm}
  \includegraphics[width=1.0\linewidth]{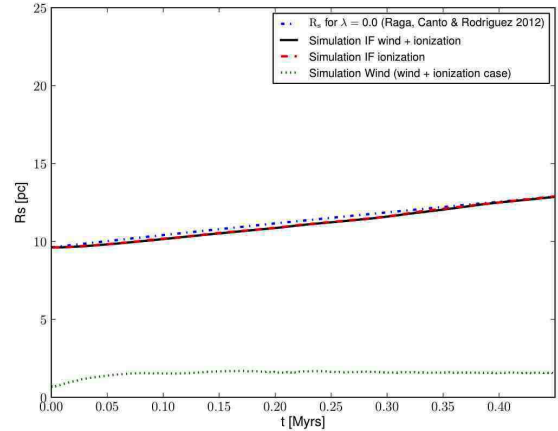}
  \caption{Evolution of the Ionization front in the wind+ionization simulation (black solid line) and in the ionization only case (red dashed line). The dashed dotted line is the result of the integration of Equation 25 from \citet{2012RMxAA..48..199R} for $\lambda=0$,  normalized  to the same starting values as given by the simulations.}
\label{O7_ionall}  
\end{figure}

The position of the ionization front is quite similar in the ionization-only run and the dual feedback run, with the ionization-only run reaching slightly higher densities at the front position. The cold material, outside the dense shell, does not feel the impact of the momentum input. Figure~\ref{uds_rho_r} also shows the rarefaction wave behind the isothermal shock front which is remarkably similar in both runs with and without winds but including ionization. 
\citet{2012RMxAA..48..199R}  present solutions for the radius of an H\,{\sc ii}~region for different values of a dimensionless parameter $\lambda$, which accounts for the relative importance of a stellar wind.  We compare the evolution of the ionization fronts in our simulations to the solution for the radius of an H\,{\sc ii}~region presented in their paper.
We find that the position of the ionization front is very similar in both our feedback cases involving ionizing radiation and the front evolution agrees with the result of the numerical integration of Equation~25 from \citet{2012RMxAA..48..199R} for their dimensionless parameter $\lambda = 0$ (see Figure~\ref{O7_ionall}). In their paper, $\lambda = 0$ describes the case of a wind-less  H\,{\sc ii}~region. 
These first test simulations already demonstrate the limited effects of wind-blown bubbles on the surroundings compared with ionization.
  \begin{figure*}[!p]
 \hspace{-2.0em}
\includegraphics[width =1. \linewidth]{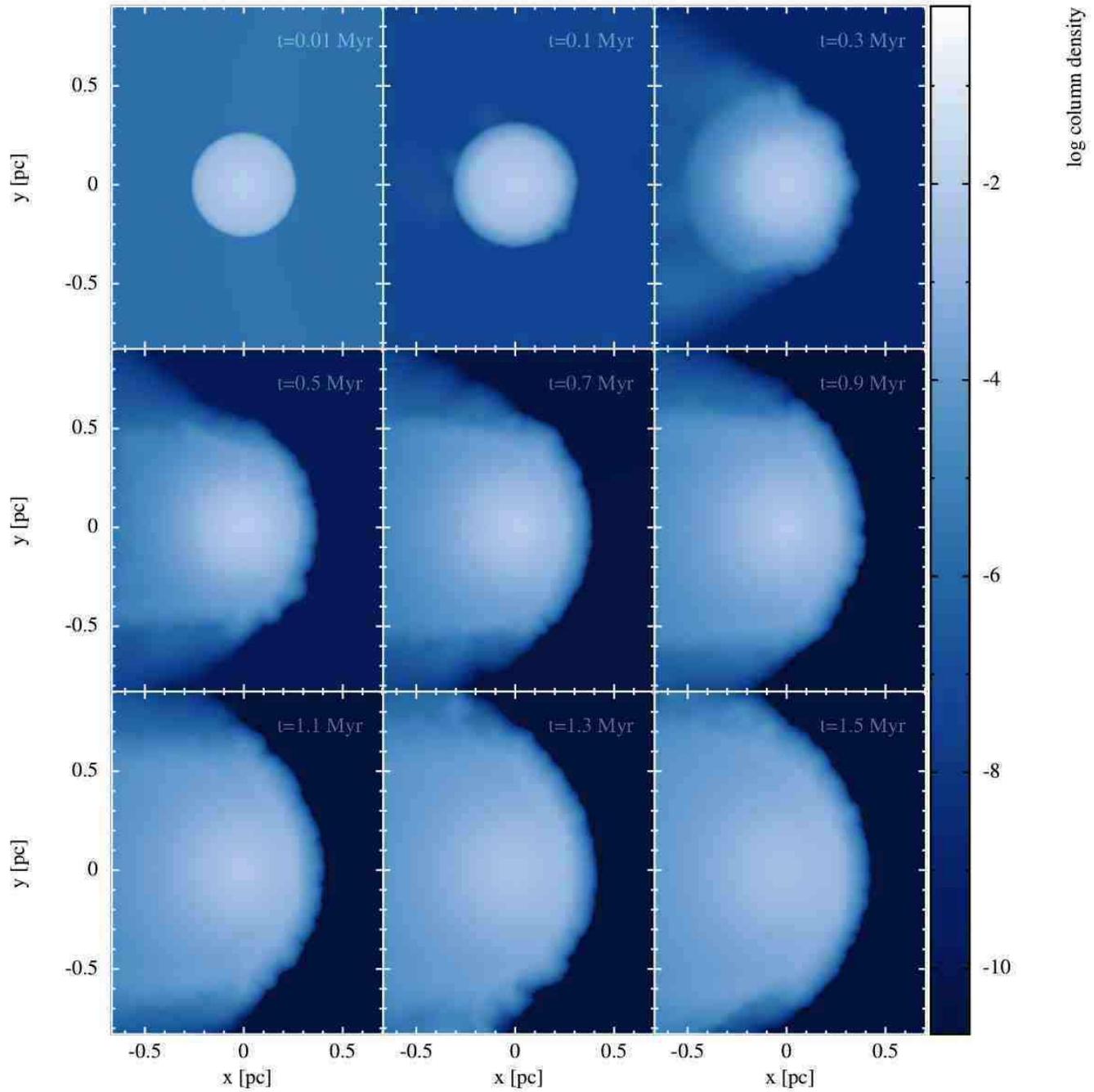}
\caption{Time  evolution of the impact of the momentum transfer on a cold core. The color bar represents the integrated density along the z-axis in $\rm{g/cm^{2}}$ }
\label{windcore}      
\end{figure*} 

%%%%%%%%%%%%% BES %%%%%%%%%%%%%%%%%%%%%%%%%%%%%%%%%%%%%%%%%%%%%

\begin{figure}[!h] 
\includegraphics[width =1.1 \linewidth]{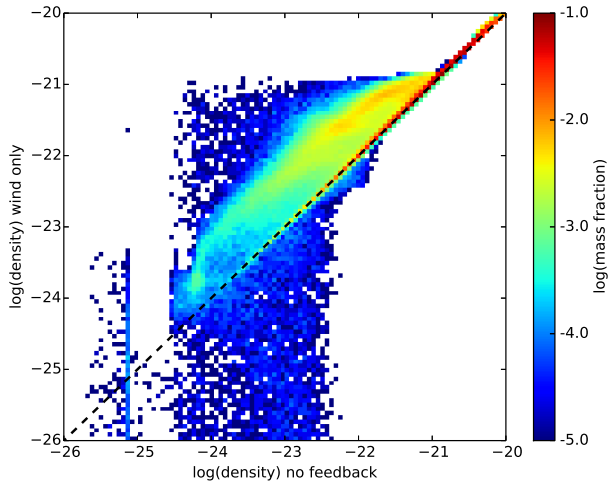}
\caption{ 2D histogram for the density distribution in the winds-only (y-axis) case and for the case without feedback (x-axis) at $t=1.2$~Myr. The color bar shows the mass contained in the bins.  The black dotted line represents the points where the density in the two runs are the same}
\label{we6rhonof}      
\end{figure} 
 
%%%%%%%%%%%%%%%%%%%%%%%%%%%%%%%%%%%%%%%%%%%%%%%%%%%%%%%%%%%

 \begin{figure*}[!p]
\hspace{-2.em}
\includegraphics[width =1. \linewidth]{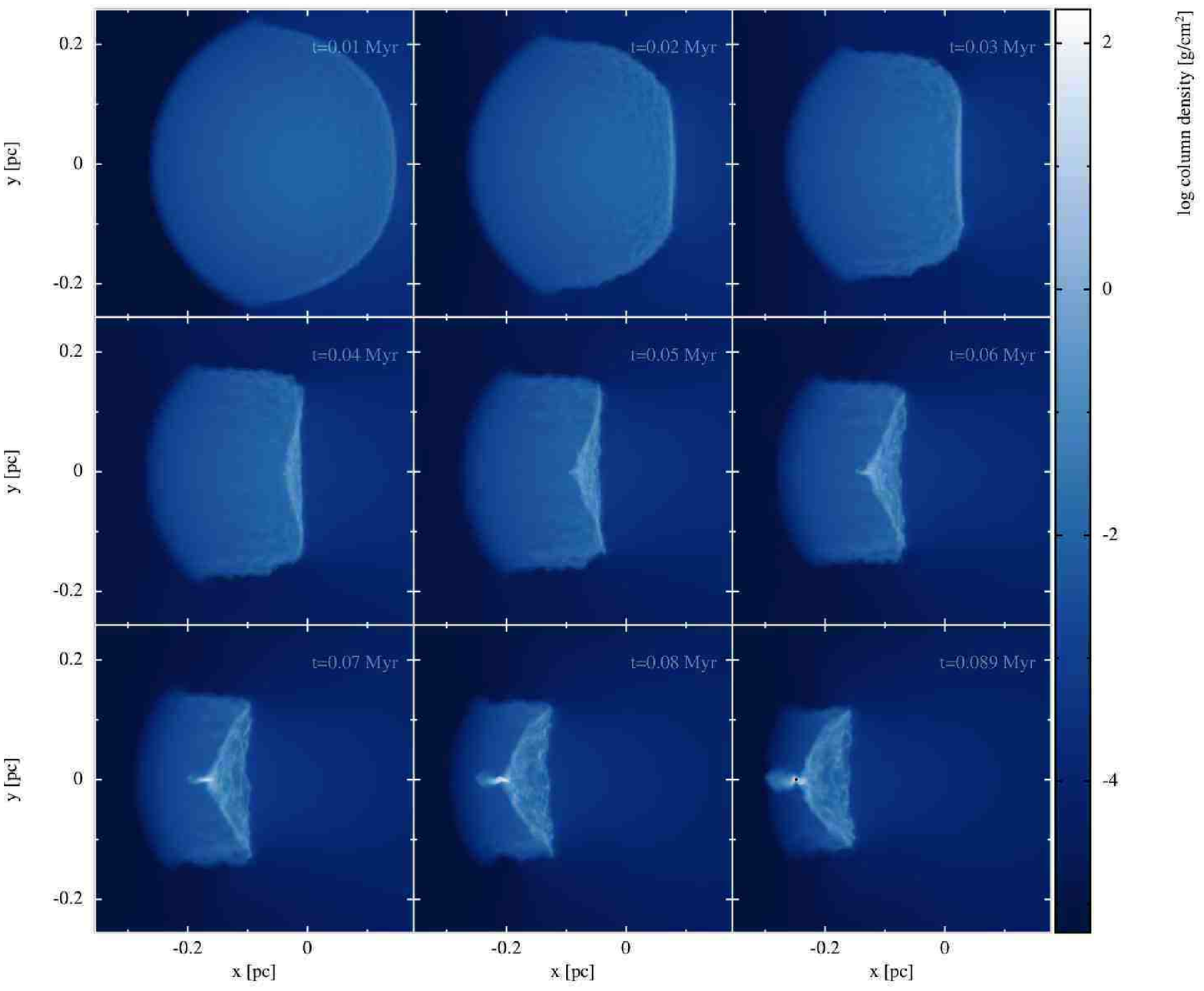}
\caption{Time  evolution of the combined impact of the momentum transfer and the ionizing radiation on a cold core. The color bar represents the integrated density along the z-axis in $\rm{g/cm^{2}}$ }
\label{BESwii}      
\end{figure*} 

\begin{figure*}[ht]
\centering
\begin{minipage}{\textwidth}
  \centering
  \includegraphics[width=0.9\linewidth]{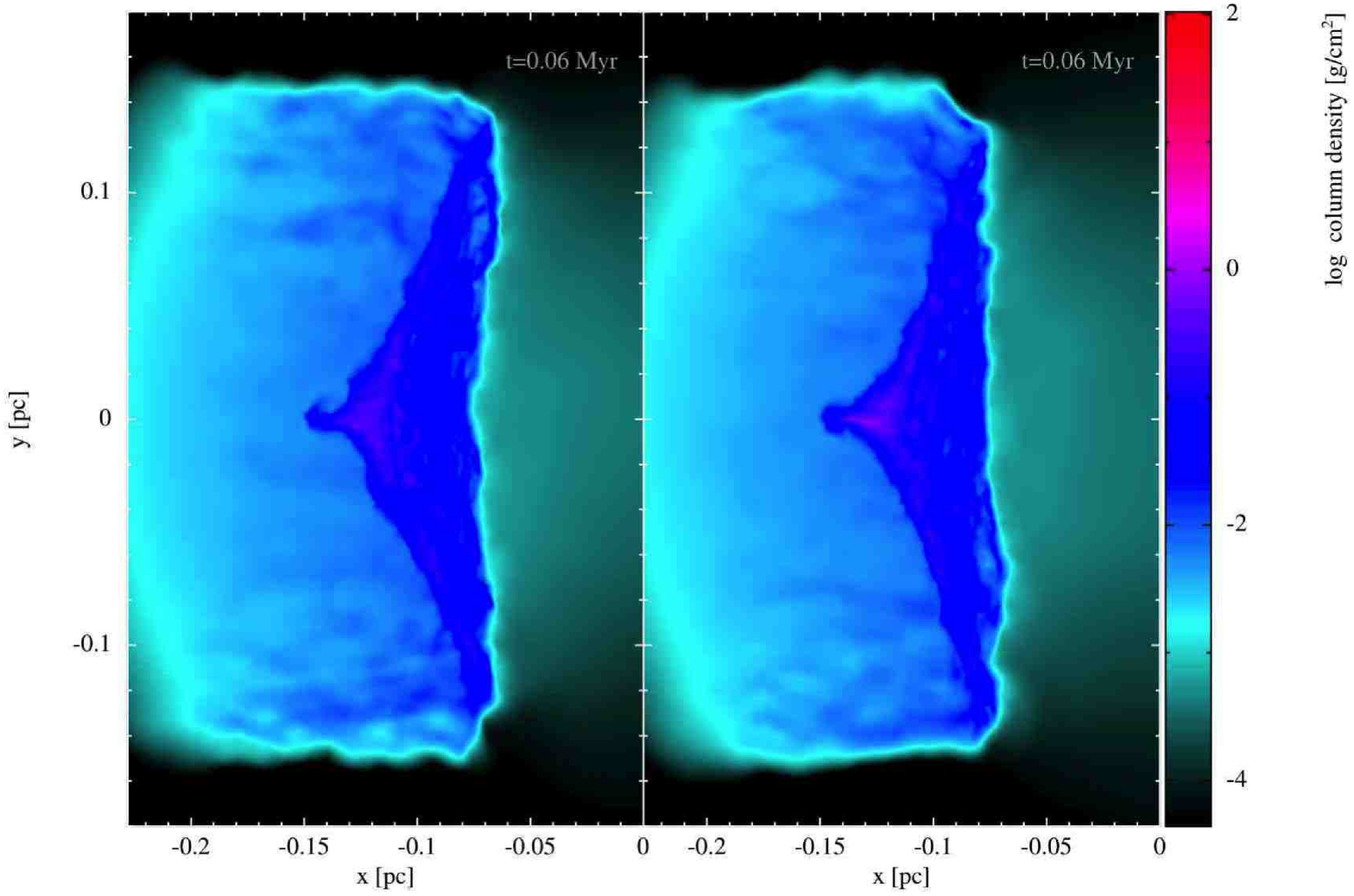}
\end{minipage}%
\newline
\begin{minipage}{\textwidth}
  \centering
  \includegraphics[width=0.9\linewidth]{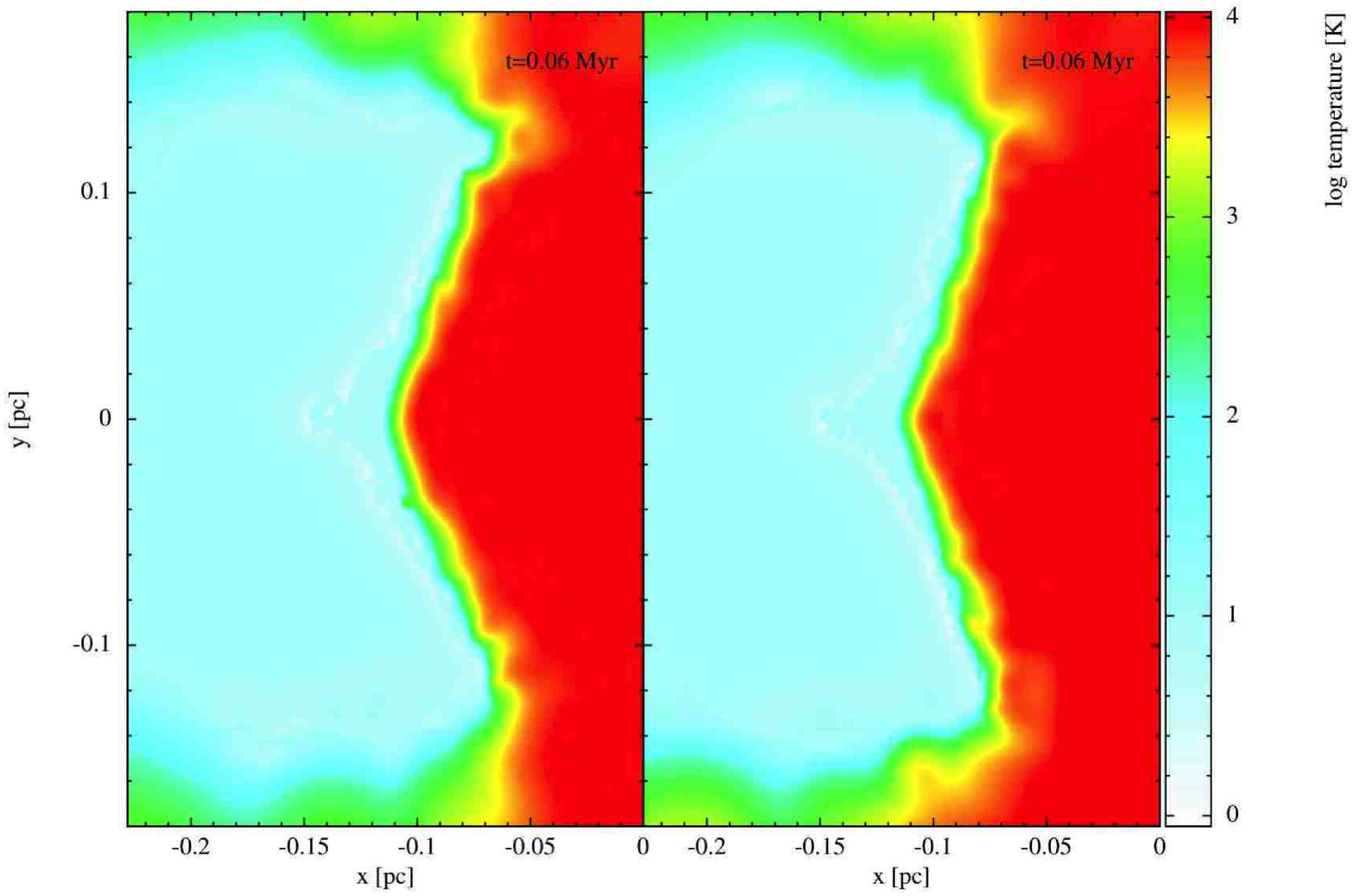}
  \caption{Snapshot of the column density (top row) and of the temperature in a slab through the center of the dense core at $z=0$ (bottom row)  for the momentum wind and ionizing radiation case (left panels) and for the ionizing radiation only case (right panels) at the same time $t=0.06\,{\rm Myr}$  }
\label{wii_coreB_t06}  
 \end{minipage}
\end{figure*}

\begin{figure*}[ht]
\centering
\begin{minipage}{\textwidth}
  \centering
  \includegraphics[width=0.9\linewidth]{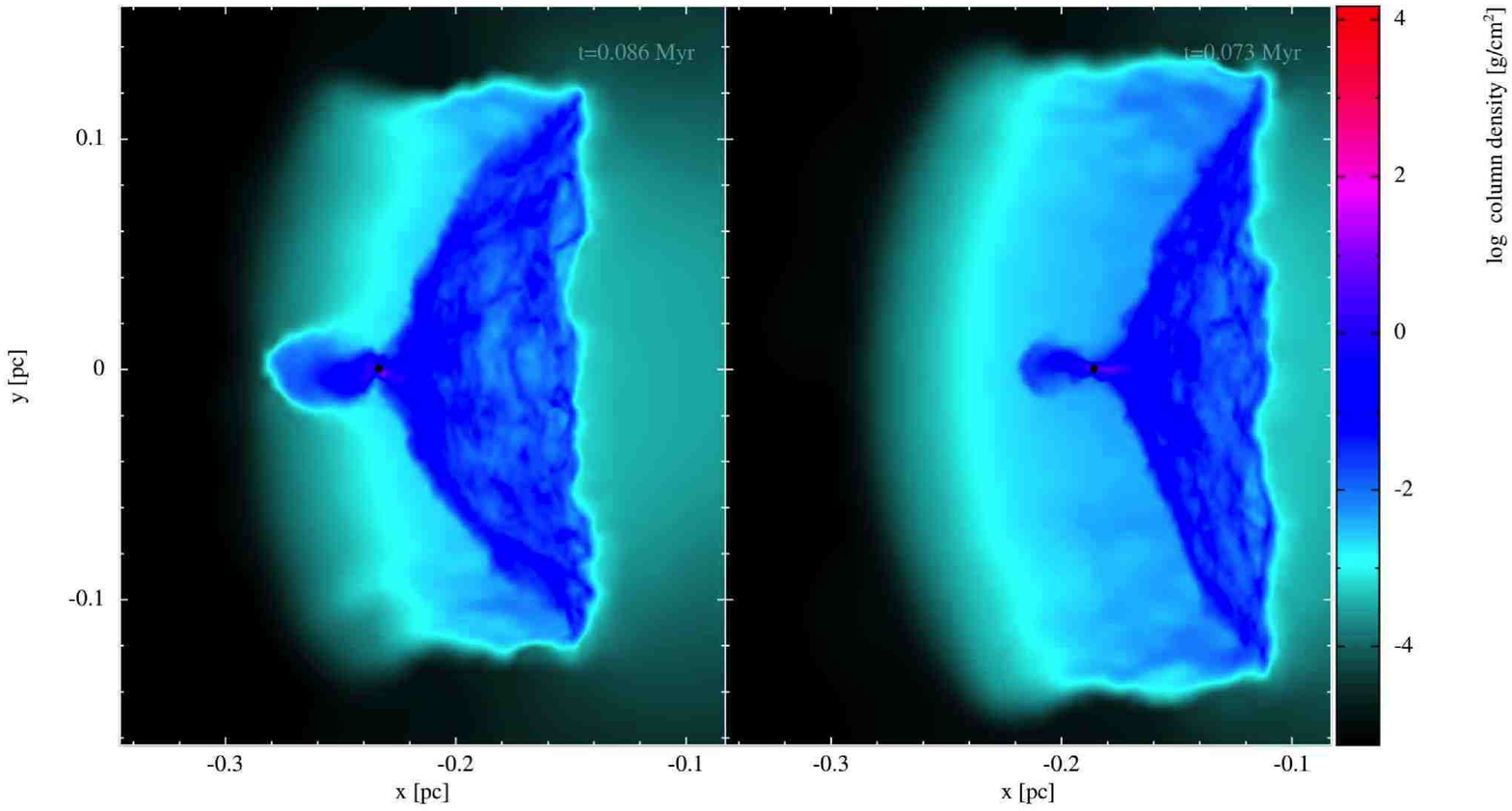}
\end{minipage}%
\newline
\begin{minipage}{\textwidth}
  \centering
  \includegraphics[width=0.9\linewidth]{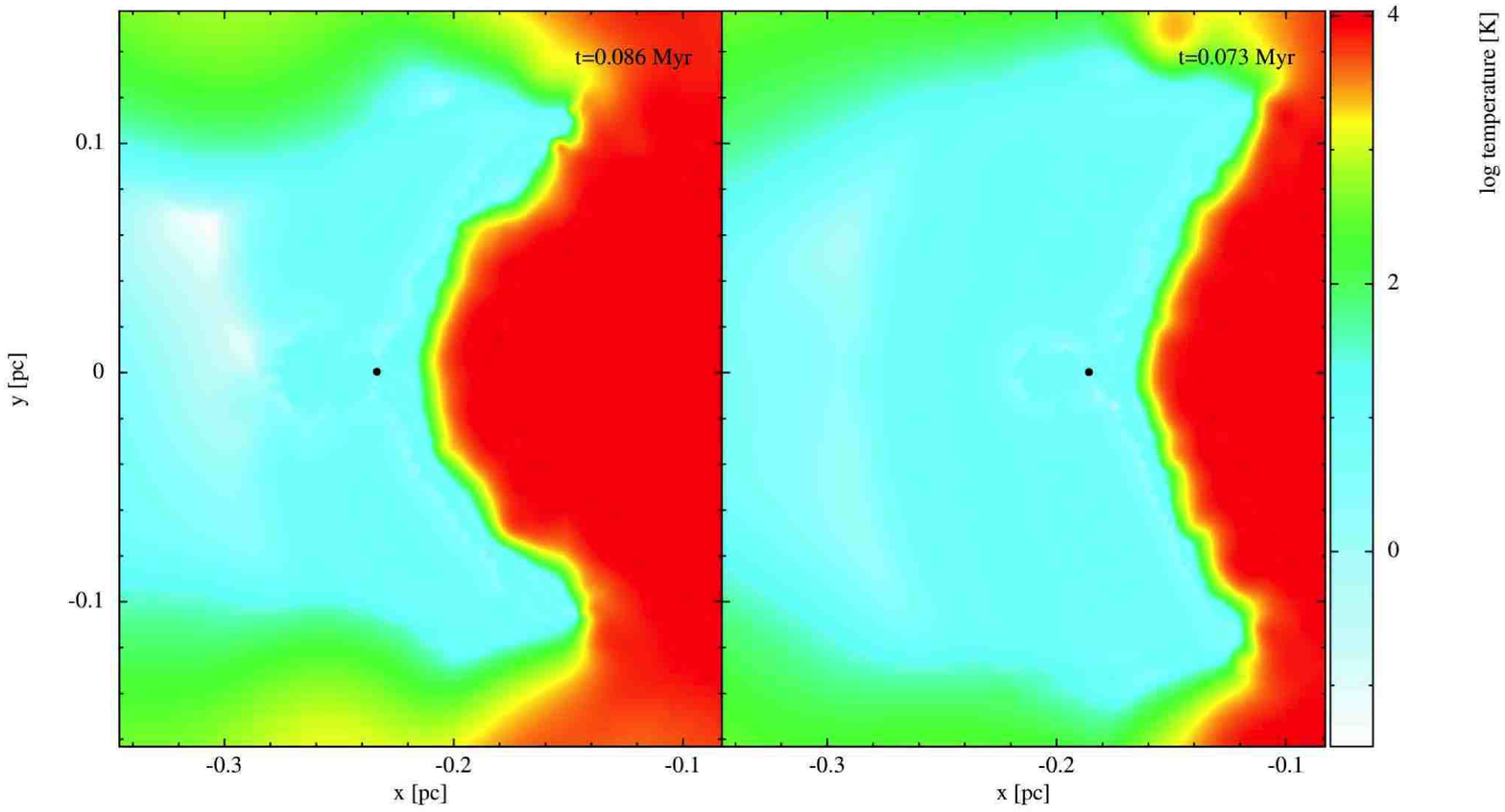}
  \caption{Snapshot of the column density (top row) and of the temperature in a slab through the center of the dense core (bottom row)  for the momentum wind and ionizing radiation case (left panels) and for the ionizing radiation only case (right panels).  The sink particles are represented by black dots. }
\label{wii_sink_coreB}  
 \end{minipage}
\end{figure*}

 \section{Effect on a Self-Gravitating Core}
 \label{sec:bes}
 
 \subsection{Initial Conditions}\label{subsec:besinicond}

We apply our numerical scheme to a self-gravitating core. We assume a dense core excavated from its molecular environment which finds itself exposed to the feedback from a massive star. 
The core is modeled as a subcritical isothermal Bonnor-Ebert sphere (BES) profile with a dimensionless boundary radius $\xi_B= 4.0$ \citep[e.g.][]{2009ApJ...695.1308B}. The temperature of the core is $T = 10\,\rm{K} $ and the isothermal sound speed is $c_{s} = 0.2\, \rm{km\,s^{-1}}$. Its mass is set to $M_{_{\rm CORE}}=4 \, \rm{M}_{\odot }$. The initial central number density is $n_0 = 6\times10^{3}\, \rm{cm}^{-3}$ and the core radius amounts to $R_{_{\rm CORE}} = 0.25 \,\rm{pc}$.  The BES is embedded in a cold uniform density medium ($T = 10 \,\rm{K}$ and $n_{_{\rm MED}} = 0.05 \, \rm{cm}^{-3}$). All SPH particles are drawn from initially settled glass-like distributions to minimize numerical noise.

We use a barotropic equation of state: 

\begin{equation}
\label{EQN:BAROTROPIC}
P=c_s^2\rho\left\{1+\left(\frac{\rho}{\rho_{_{\rm CRIT}}}\right)^{\gamma -1}\right\}\,,
\end{equation}

\noindent where $P$ is the thermal pressure of the gas, $\rho$ is the gas density, $\rho_{_{\rm CRIT}} =10^{-13}\, \rm{g\,cm}^{-3}$ is the critical density above which the gas becomes approximately adiabatic, $c_s = 0.2\, \rm{km\,s^{-1}}$ for molecular hydrogen at $T = 10\,\rm{K}$  and $\gamma = 5/3$ is the ratio of specific heats. This value of $\gamma$ is justified as we treat $T=10\,K$ where the rotational degrees of freedom for $\rm{H}_2$ are not highly excited.
Local density peaks with $\rho_{_{\rm PEAK}} > \rho_{_{\rm SINK}} = 10^{-11}\,\rm{g\, cm}^{-3}$ are replaced by sink particles which then accrete mass using the newly developed algorithm of \citet{2013MNRAS.430.3261H} which regulates the accretion of matter onto a sink and redistributes the angular momentum of the accreted material to the surrounding gas.
We use $5\times10^5$ particles to model the BES, resulting in a particle mass  $m_{_{\rm PART}}=8\times10^{-6}\, \rm{M}_{\odot }$. The minimum Jeans mass corresponding to a critical density $\rho_{_{\rm CRIT}} = 10^{-13}\, \rm{g\,cm}^{-3}$ at a temperature $T = 10\,\rm{K}$ is $M_J=3\times10^{-3}\, \rm{M}_{\odot }$ and is therefore always resolved \citep[][]{1997MNRAS.288.1060B}, as $2m_{_{\rm PART}} N_{_{\rm NEIGH}} = 8\times10^{-4}\, \rm{M}_{\odot}$ and $N_{_{\rm NEIGH}} = 50$ being the number of SPH neighbors. 
The core is then exposed to three different types of feedback from a source placed at a distance of $d_s=3\,\rm{pc}$ from the core center. Since $d_s \gg R_{_{\rm CORE}} $, the stellar feedback is impinging in an almost plane parallel fashion on the core.

\subsection{Momentum Winds Only} \label{subset:besmomonly}

To examine the impact of the momentum transfer on the core, we used our fiducial values for the stellar mass loss rate $\dot{M}_{_{\rm WIND}}= 10^{-6}\,\rm{M}_{\odot}\,\rm{yr^{-1}}$ and the wind terminal velocity $v_{_{\rm WIND}}= 2000 \,\rm{km\,s^{-1}}$. 
 Figure~\ref{windcore} shows a time sequence of the evolution of the core. The cold material is slowly ablated from the front side of the core and redirected to the sides. The material at the back, which is shielded from the wind, expands into the lower pressure environment. Over time the front material at intermediate densities is slowly compressed. However the wind has very little effect on the densest inner region of the core. The extra compression is not enough to induce gravitational collapse.  
After $\sim 1 \,\rm{Myr}$, which also corresponds to the free-fall time in the center, the densest material starts to be dispersed by the expansion of the core which then quickly dissolves.

Both the dispersive and the compressive effects are illustrated in  Figure~\ref{we6rhonof}, which shows a two dimensional histogram comparing the density in the fiducial wind simulation (y-axis) and the density in the no-feedback case (x-axis) at $t=1.2$~Myr, when the highest density is reached in the center. The black dotted line shows equal densities. It represents gas which density is not affected by feedback.
Filled histogram bins above it represent material that has an increased density in the wind-only run, while those below represent material that has a lower density compared to the no-feedback run.
 Figure~\ref{we6rhonof} shows that the momentum transfer mostly affects the low and intermediate-density material at the front edge of the core. The largest spread around the $x = y$ line is seen for densities  between $10^{-24}$ and $10^{-21}\, \rm{g\,cm}^{-3}$. Most of the mass is above the line indicating the compressive effect of the wind. A slight density increase can also be seen for higher densities  $\geq10^{-21}\, \rm{g\,cm}^{-3}$ but the impact of the wind is rather modest.

\subsection{Combining the Momentum Transfer and the Ionizing Radiation} \label{subsec:wind+ion}

We now look at the combined effects of the ionizing radiation \emph{and} the momentum wind from our fiducial feedback source on our BES. The values for the stellar mass loss and the terminal wind velocity are the same as  above. The ionizing photon rate is set to $\dot{N}_{_{\rm LyC}}=10^{49}\,\rm{s^{-1}}$. The core is located well within the source's initial Str\"omgren-radius and finds itself embedded in a warm ($T=10^4\,\rm{K}$) environment. 

\begin{figure}[!h]
\includegraphics[width = 1.\linewidth]{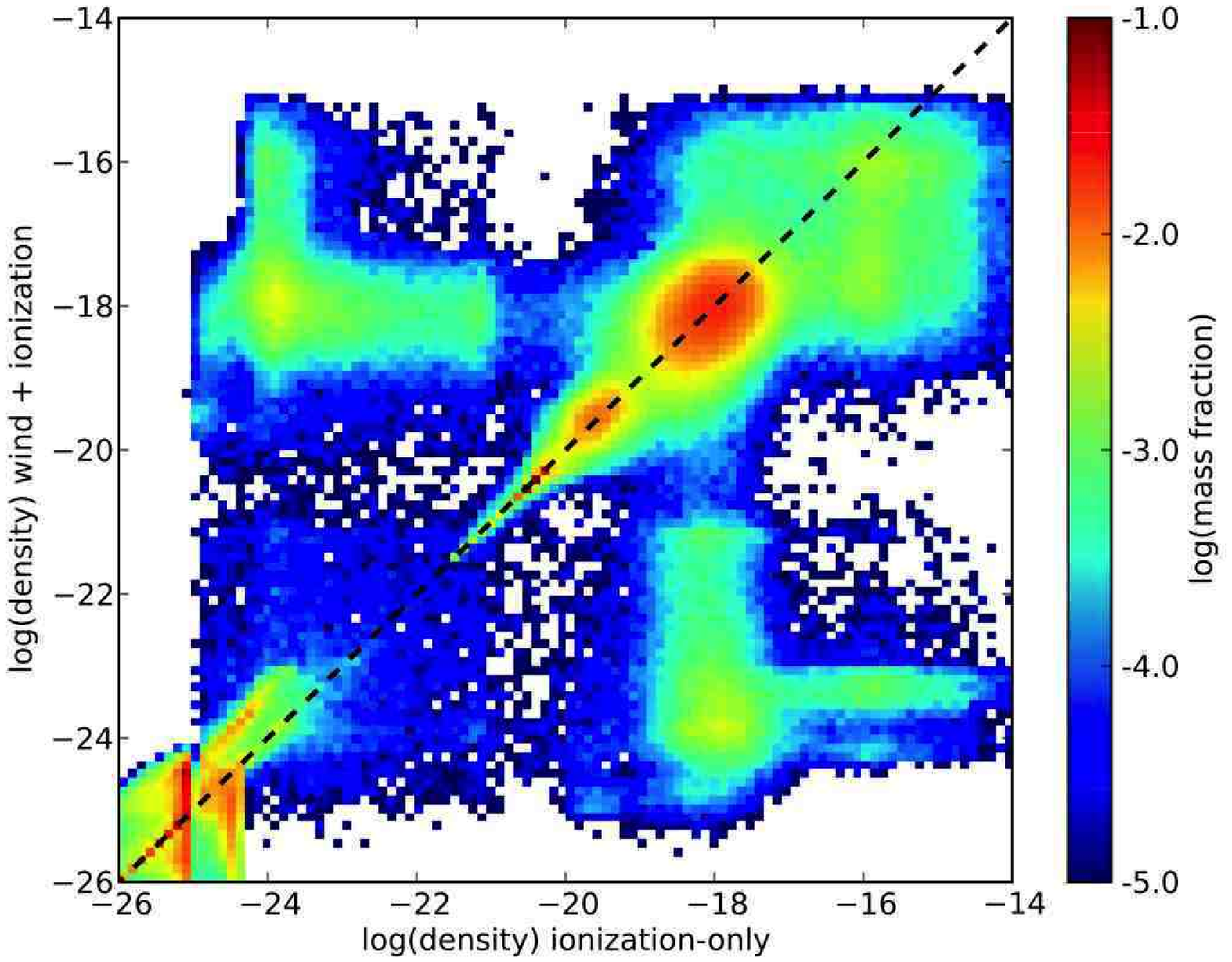}
\caption{ 2D histogram for the density distribution in the wind and ionization case (y-axis) case and for the case with ionization only (x-axis) at $t=0.07$~Myr. The color bar shows the mass contained in the bins. The black dotted line represents the points where the density in the two runs are the same. }
\label{O7_wion_c018_join}
\end{figure} 
\begin{figure}[!h]
  \subfigure{\includegraphics[width=1.\linewidth]{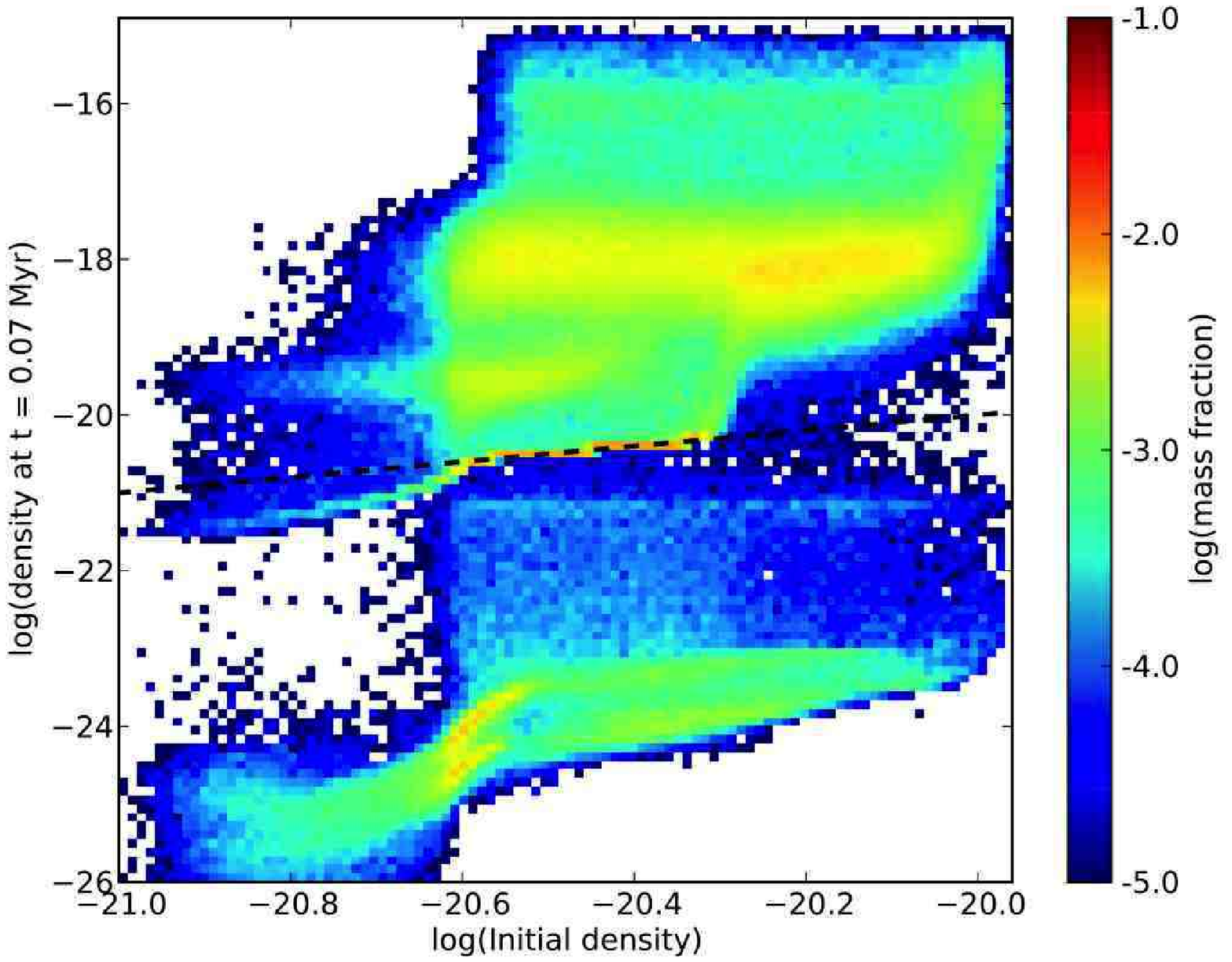}}
\caption{Two-dimensional histogram for the fiducial O7.5 star showing the densities at $t=0.07\,\rm{Myr}$ on the vertical axis as a function of the initial densities of the core. The color bar shows the mass contained in the bins.}
 \label{O7_hist}
\end{figure}

Figure~\ref{BESwii} shows a time sequence of column density plots for the evolution of the core under the impact of the combined feedback mechanisms. 
The ionization front compresses the illuminated front of the core while the sides are compressed by the pressurized ambient medium. 
The material at the edge is photo-evaporated. The back of the core is initially shielded from the ionizing radiation  by the denser core but is quickly filled by low-density ionized gas from the side.  A shell of swept-up gas builds up a the front of the cloud (e.g. panel~3 at $t=0.03\,\rm{Myr}$)
The momentum transfer through the evaporation of the illuminated front is strong enough to displace the  core; panel~4 at $t=0.04\,\rm{Myr}$ shows that the initial center of the core is pushed in negative x-direction. 
After $\sim 0.05 \rm{Myr}$, the swept-up shell  
contracts laterally due to the outside pressure of the ionized gas.   
Through the combined effect of the movement in x-direction and the contraction towards the densest part
due to the compression in the y and z-directions, the initially spherical core forms a dense elongated structure. At $t \approx 0.08 \rm{Myr}$, the densest region of the filament  collapses to form a sink particle.  Similar to the low ionizing flux runs described in \cite{2009A&A...497..649B}, star formation first appears  ahead of the ionizing front towards the center of the core.  

  \begin{figure}[h]
  \includegraphics[width = \linewidth]{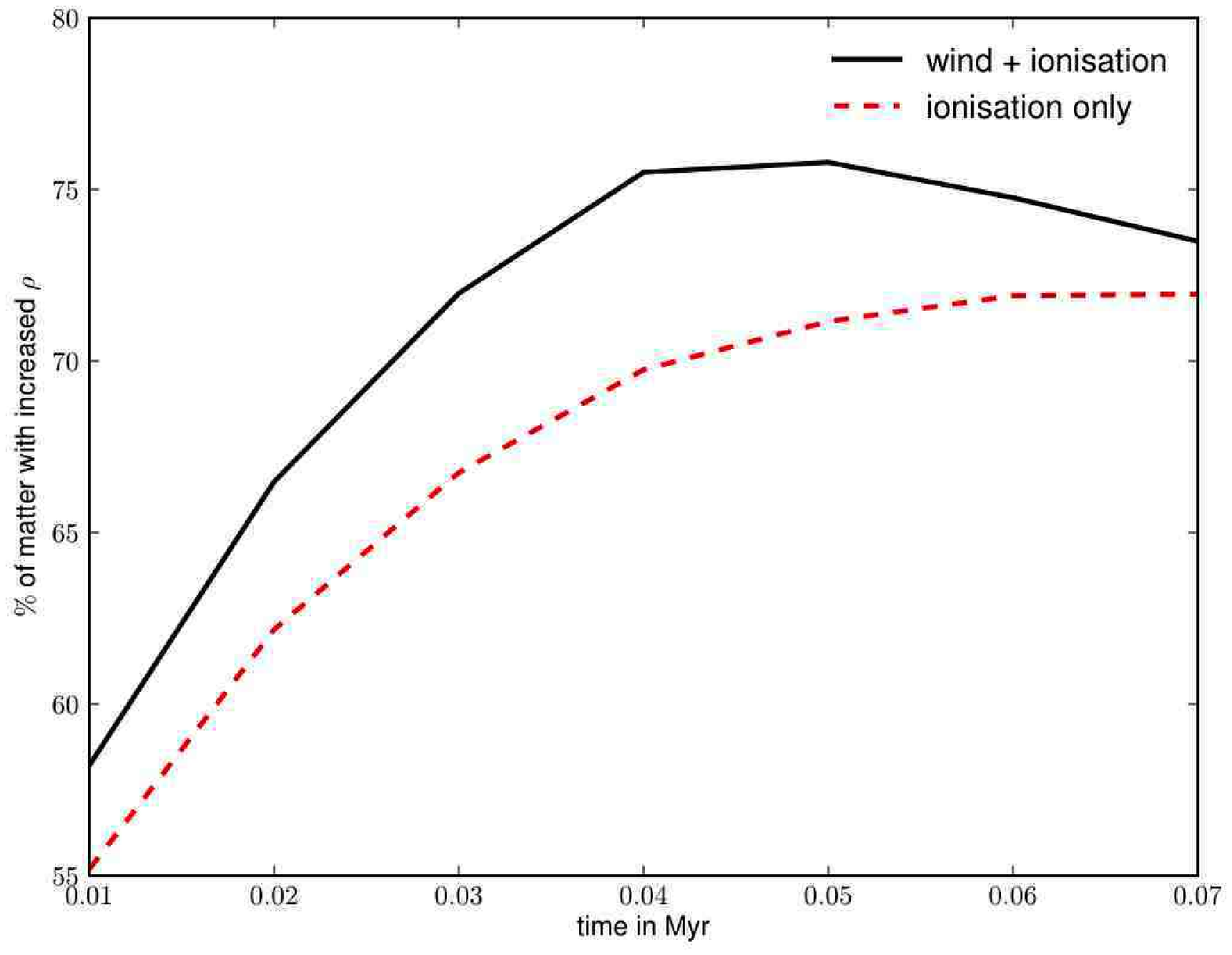}
\caption{Percentage of particles with increased density since $t_0$ for the wind and ionization case (solid black line) and the ionization-only case (dashed red line) as a function of time }
\label{comp_wii}  
\end{figure}

 \begin{figure}[h]
\includegraphics[width = \linewidth]{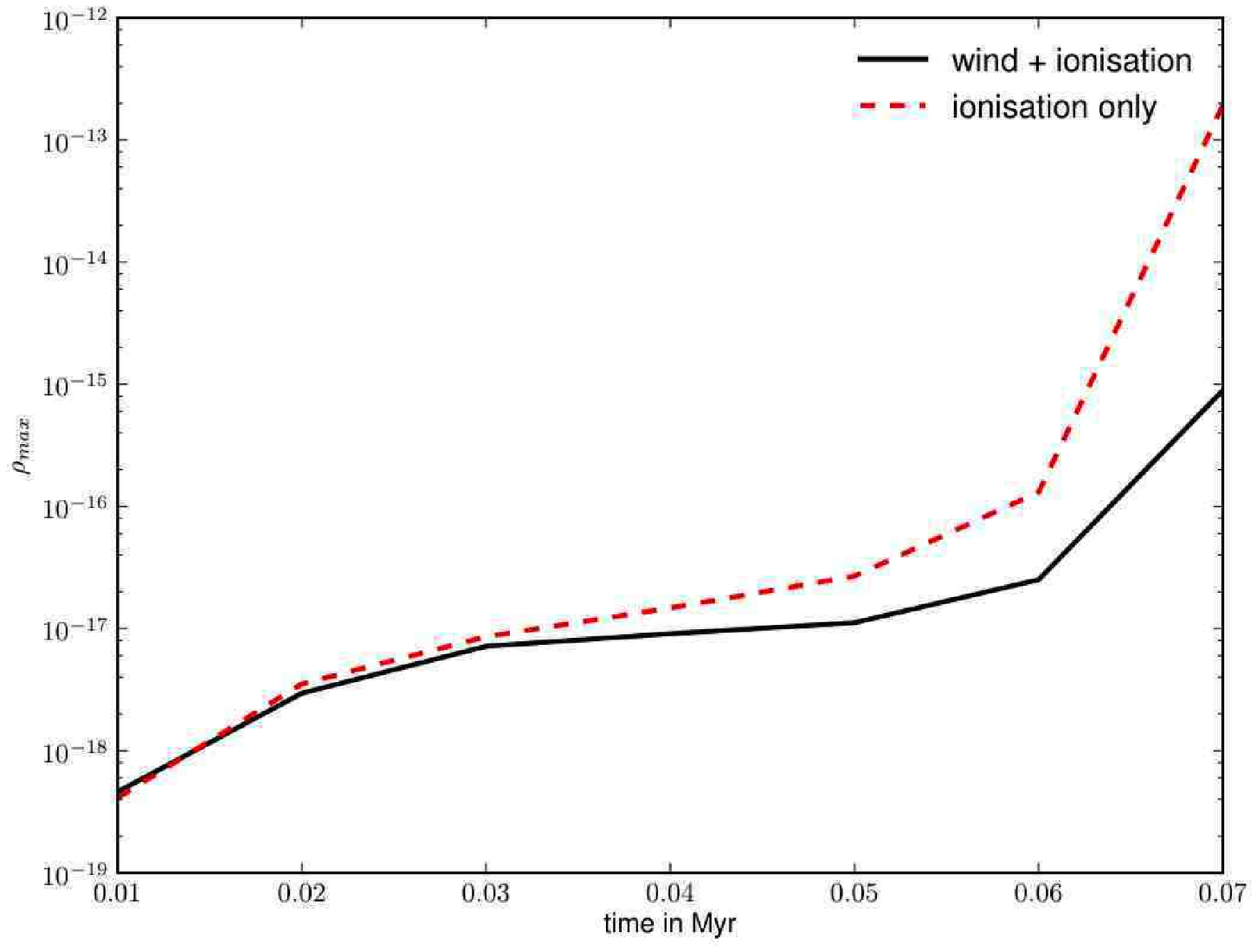}
\caption{Time evolution of the maximum of the density of the ionized self-gravitating core for the fiducial feedback values. Ionizing radiation only (dashed red);  momentum wind and ionizing radiation (solid black). }
\label{rhomax_wii_fid}       
\end{figure} 

Figure~\ref{wii_coreB_t06} shows snapshots of  the combined feedback run (left panels) and the ionization-only run (right panels) at a same time $t=0.06\,{\rm Myr}$. The top row displays the column density and the bottom row shows the temperature. The  appearance of the cold gas is remarkably similar in both cases. The ionization fronts are at the same location (see temperatures in the bottom row).    
 The structure appears slightly less compressed in the combined-feedback case. The densities at the tip of the converging filament structure are a little higher in the ionization-only case. 
In Figure~\ref{wii_sink_coreB} we compare the combined feedback run (left panel) with the ionization-only run (right panel) at a time just after sink formation.  The left panel, corresponding to the dual-feedback run,  is at a slightly later time than the right panel indicating that the addition of the momentum wind leads to a small delay in sink particle formation.
The first sink particle is formed after $\sim 0.086\,\rm{Myr}$ in the dual-feedback case, a  bit later than in the ionization-only run where the first sink appears at  $\sim 0.073\,\rm{Myr}$. The overall appearance of the core however is still quite similar  in both cases.

Figure~\ref{O7_wion_c018_join} shows the 2D histogram of particle densities $\rho_{_{\rm DUAL}}$ (x-axis) and  $\rho_{_{\rm ION}}$ (y-axis) in the wind and ionization simulation  and the ionization-only simulation respectively.  It shows the distribution of particles in density space at $t=0.07$~Myr, a time just before sink formation in the ionization-only case. 
Most of the particles have densities around $\sim 10^{-18}\, \rm{g\,cm}^{-3}$. They are part of the dense filament and the shell like structure at the front edge of the core. The distribution in the histogram appears almost symmetric around the black dotted line. {This shows that the density distribution is very similar in both simulations. The spread around the black dotted line shows that the particles contributing to the different density phases are not entirely the same.  The area above the black dotted line shows gas  with $\rho_{_{\rm DUAL}} > \rho_{_{\rm ION}}$, for which the momentum wind lead to an increase in density while the area below indicates gas with $\rho_{_{\rm DUAL}} < \rho_{_{\rm ION}}$. 
The momentum wind has a dual impact. It both compresses and disperses the gas.}

{
Some of the particles making up the core in the combined case  are found in the low-density regime in the ionization-only case, and vice-versa. This indicates that, although there are few differences between the two feedback runs,  the material contributing to the formation of denser structures can be different when including the effect of the winds. This could be of  some meaning for the
microphysics and chemistry of gas mixing as the winds can contribute to the localized redistribution and reshuffling of gas. } 

In Figure~\ref{O7_hist} we compare the densities at $t=0.07\,\rm{Myr}$ to the initial densities of the same material in the core. Approximately $73\%$ 
of the particles have a higher density at $t=0.07\,\rm{Myr}$ in the combined-feedback run for $71\%$
in the ionization-only case.  However, $49 \%$ of the particles have a higher density in the combined-feedback run than in the corresponding ionization run.  Although in the dual-feedback run slightly more gas has increased its density since $t_0$ (Figure~\ref{comp_wii}), the ionization run appears to have the highest densities (see Figure~\ref{rhomax_wii_fid}). 
 {Effectively the wind, through its ram pressure,  slightly increases the density of the ionized gas between the core and the source. This leads to less-ionized particles in the core since the recombination rate depends on  $n_{\rm e}^2$.  Where $n_{\rm e}$ is the electron number density, which approximately equals the ionized gas number density for hydrogen.   
 At the same time part of the ionized gas at the sides of the cloud is blown away by the wind,
which results in the core being compressed a little slower in the combined case. Overall the core contains more neutral gas but is also less compressed in the dual-feedback case. This leads to delayed star formation.
This effect is very small, however, as the position of the ionization fronts and the overall density distribution are very similar in both feedback runs (see Figure~\ref{wii_coreB_t06}).}

\subsubsection{Impact of a B0 Star } 

To study the impact of a fainter massive star, we expose the core to the ionizing radiation and the wind momentum from a B0 star with much weaker winds. We adopt values from \cite{2006MNRAS.367..763S} in his census of the massive star in the Carina Nebula. We use $\dot{M}_{_{\rm WIND}}= 3\times10^{-7}\,\rm{M}_{\odot}\,\rm{yr^{-1}}$ and $v_{_{\rm WIND}}= 1180 \,\rm{km\,s^{-1}}$ for the mass loss rate and the terminal wind velocity and an ionizing photon rate of $\dot{N}_{_{\rm LyC}}=1.9\times10^{48}\,\rm{s^{-1}}$. 

 \begin{figure}[h]
\includegraphics[width = 1.0\linewidth]{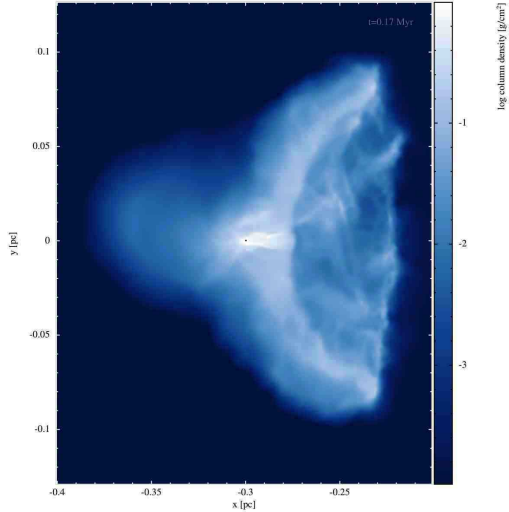}
\caption{ {Snapshot of the column density showing the combined impact of the wind and ionizing radiation from a B0 like star at a time $t = 0.17\,{\rm Myr}$ just after sink formation. }}
\label{B0_017}      
\end{figure} 
The ionization front advances slower than in our fiducial case. The front appears more extended and fuzzy (see Figure~\ref{B0_017}). A similar behavior to the fiducial case is observed. The material is swept up in a dense front that contracts and collapses towards the symmetry axis. The morphology of the core resembles the concave shape (with respect to the feedback source) described in the O7.5-star case.    
The first sink particle is formed significantly later than in our fiducial case, at  $t_*\approx 0.17\,\rm{Myr}$ in the dual feedback run and at  $t_*\approx 0.18\,\rm{Myr}$ in the ionization-only run.  In this case the momentum wind leads to slightly earlier star formation.

\subsubsection{Impact of an O3 Star} 

We also selected a more powerful source at the upper end of the massive star range.  We use values from \cite{2006MNRAS.367..763S} for an O3 star with a mass loss rate, a terminal wind velocity and ionizing photon rate of  $\dot{M}_{_{\rm WIND}}= 1.3\times10^{-5}\,\rm{M}_{\odot}\,\rm{yr^{-1}}$, $v_{_{\rm WIND}}= 3160 \,\rm{km\,s^{-1}}$ and $\dot{N}_{_{\rm LyC}}=6\times10^{49}\,\rm{s^{-1}}$.  

 \begin{figure}[h]
\includegraphics[width = 1.0\linewidth]{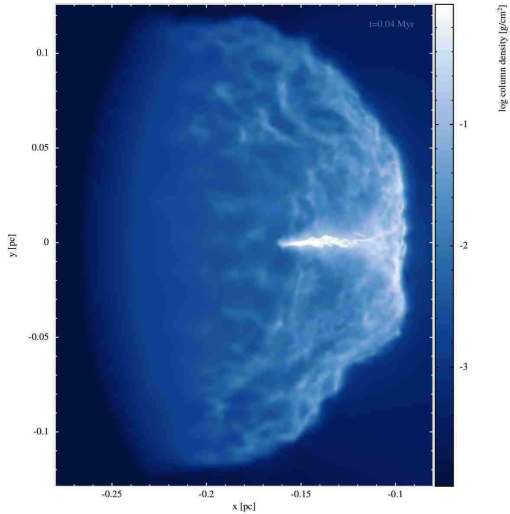}
\caption{ {Snapshot of the column density showing the combined impact of the wind and ionizing radiation from an O3 like star at a time of greatest compression ($t = 0.04\,{\rm Myr}$). }}
\label{O3_004}      
\end{figure} 
The evolution of the morphology of the core in this case differs from the ones we obtain with the less massive stars. 
Instead of the concave form described above,  the core evolves into a convex shape  (see Figure~\ref{O3_004}). The front is being accelerated inside the core and the less dense structures have a higher velocity than the denser ones along the symmetry axis. {The material  converges towards the symmetry axis due to the outer pressure of the ionized gas. }
A central filament forms but the material is evaporated, ionized and dispersed before it can fragment.  No sink particle is formed.

%%%%%%%%%%%%%%%%%%%%%%%%%%%%%%%%%%%%%%%%%%%%%%%%%%%%%%%%%%%

\section{Summary and Discussion}
 \label{sec:conc}

We present the implementation of a new momentum wind scheme for the SPH code SEREN and study the impact of the momentum transfer from stellar wind ejecta on the surrounding molecular and ionized density distribution. This scheme is particularly suitable for  modeling massive star feedback in simulations of star formation in a cluster environment. It can be used in conjunction with the ionization scheme described in \cite{2009A&A...497..649B}.

We look at the impact of momentum winds and ionization from a massive star on a uniform density environment. {We use a spherical uniform density cloud  of density ${n}_{\rm c} = 30\,{\rm cm}^{-3}$, of radius $R_{\rm c} = 16\,{\rm pc}$ and a  temperature $T =10$~K. The feedback source is located at the center.} We use values for the wind mass loss $ \dot{M}= 10^{-6}\,{\rm M}_{\odot}\, {\rm yr}^{-1}$, the wind velocity $v_{_{\rm WIND}} = 2000\,{\rm km}\, {\rm s}^{-1}$ and the ionizing photon rate  ${ N}_{{\rm Lyc}} = 10^{49}\,{\rm s}^{-1}$, close to those for an O7.5-star as cataloged by  \citet{2006MNRAS.367..763S} in his census of massive stars in the Carina Nebula.
We find that:

i) In a cold molecular environment, the pure transfer of momentum from the stellar wind is able to sweep up and compress the gas. It never reaches the gas densities which are obtained in the ionization runs. This makes momentum winds much less efficient than ionizing UV radiation in compressing cold gas and eventually triggering star formation.

ii) During the main-sequence life of a massive star,  stellar winds do not act on their own but in combination with the ionizing radiation.  In this combined case, the ionizing radiation appears to be the main agent in shaping and compressing  the cold gas. The momentum wind affects only the inner most part of the ionized region.  {This leads to  
an H\,{\sc ii}-region with a small hole around the feedback source. }

iii) {We apply the wind and ionization feedback o a self-gravitating core  with a Bonnor-Ebert density profile. The core is modeled as a subcritical isothermal Bonnor-Ebert sphere (BES) profile with a dimensionless boundary radius $\xi_B= 4.0$ and a mass of $M_{_{\rm CORE}}=4 \, \rm{M}_{\odot }$ at a temperature of $T = 10\,\rm{K} $. The initial central number density is $n_0 = 6\times10^{3}\, \rm{cm}^{-3}$ and the core radius amounts to $R_{_{\rm CORE}} = 0.25 \,\rm{pc}$.  The BES is embedded in a cold uniform density medium ($T = 10 \,\rm{K}$ and $n_{_{\rm MED}} = 0.05 \, \rm{cm}^{-3}$). These simulations show that  the momentum wind alone has both a compressive and dispersive effect on the core. The compression does not lead to the highest densities we obtain in runs including ionization. The  compression by the momentum wind is not enough to induce gravitational collapse.}

 iv)  In the combined feedback case, the ionizing radiation is mostly responsible for compressing the core and inducing collapse.  The dense filamentary structures forming are very similar in the dual feedback case and in the ionization-only case. {However, in comparison to the ionization-only run, the dense core contains more neutral gas but is also less compressed in the dual-feedback case. }
   
v)  When exposed to different  feedback strength, the weak and intermediate-feedback runs (modeled as the effect of a B0 star and O7.5 star respectively) lead to the formation of a sink particle. For the B0 case, sink formation occurs a bit earlier in the dual feedback run, while in the O7.5 case sink formation is slightly delayed in the dual-feedback case.
For the strong feedback case, no sink particle is formed. The cold material is evaporated before it can become dense enough to undergo gravitational collapse.

The ionizing UV radiation is the main driver of  the dynamical evolution of the gas. In the case of triggered star formation, the additional momentum wind does not substantially change the outcome. It might lead to slightly earlier or delayed star formation, but it is unlikely to trigger any extra star-forming events that wouldn't happen in the ionization-only case.  Overall the contribution from the additional momentum from winds to the dynamics of molecular gas and its impact on star formation is very modest. Simulations by \cite{2014MNRAS.442..694D} of the combined effects of photoionization and momentum-driven winds on giant molecular clouds have lead to similar conclusions. Their finding that the momentum wind has little effect on the densest and most massive regions,
is confirmed by our comparison of the effects of the different types of feedback on a self-gravitating core. The overall appearance and evolution of the dense gas is almost indistinguishable in the ionization-only case and in the wind+ionization case. We argue that observations of the dynamics of cold molecular gas in cores and clumps is not likely to provide much information on the role of winds in their evolution.  But as our simulations show, the material accumulated in the denser structures and eventually involved in sink formation is partially different. This indicates that winds might contribute to the localized redistribution and mixing of gas and could thus impact the metallicity distribution in the vicinity of massive stars.

%%%%%%%%%%%%%%%%%%%%%%%%%%%%%%%%%%%%%%%%%%%%%%%%%%%%%%%%%%%%

\begin{acknowledgements}
This project is funded by the \emph{Deut\-sche For\-schungs\-ge\-mein\-schaft, DFG\/} PR 569/9-1. Additional support came from funds from the Munich Cluster of Excellence: "Origin and Structure of the Universe".
{We thank the referee for very helpful and constructive comments.}
We made use of the SPLASH software package \cite{2007PASA...24..159P} to visualize our simulations.
\end{acknowledgements}

%%%%%%%%%%%%%%%%%%%%%%%%%%%%%%%%%%%%%%%%%%%%%%%%%%%%%%%%%%%%


\begin{thebibliography}{7}

\bibitem[Arthur(2007)]{2007dmsf.book..183A} Arthur, S.~J.\ 2007, Diffuse 
Matter from Star Forming Regions to Active Galaxies, 183 

\bibitem[Bate 
\& Burkert(1997)]{1997MNRAS.288.1060B} Bate, M.~R., \& Burkert, A.\ 1997, \mnras, 288, 1060

\bibitem[Bisbas et~al.(2009)]{2009A&A...497..649B} Bisbas, T.~G., W{\"u}nsch, R., Whitworth, A.~P., \& Hubber, D.~A. 2009,
 \aap, 497, 649
  
%  \bibitem[Bisbas et al.(2011)]{2011ApJ...736..142B} Bisbas, T.~G., 
%W{\"u}nsch, R., Whitworth, A.~P., Hubber, D.~A., 
%\& Walch, S.\ 2011, \apj, 736, 142 

\bibitem[Burkert 
\& Alves(2009)]{2009ApJ...695.1308B} Burkert, A., \& Alves, J.\ 2009, \apj, 695, 1308

\bibitem[Capriotti 
\& Kozminski(2001)]{2001PASP..113..677C} Capriotti, E.~R., \& Kozminski, J.~F.\ 2001, \pasp, 113, 677

\bibitem[Castor et al.(1975)]{1975ApJ...200L.107C} Castor, J., McCray, R., 
\& Weaver, R.\ 1975, \apjl, 200, L107 
 
 \bibitem[Chu et al.(2006)]{2006ESASP.604..363C} Chu, Y.-H., Gruendl, R.~A., 
\& Guerrero, M.~A.\ 2006, The X-ray Universe 2005, 604, 363 
 
\bibitem[Dale 
\& Bonnell(2011)]{2011MNRAS.414..321D} Dale, J.~E., \& Bonnell, I.\ 2011, \mnras, 414, 321 
 
 \bibitem[Dale et al.(2013)]{2013MNRAS.436.3430D} Dale, J.~E., Ngoumou, J., 
Ercolano, B., \& Bonnell, I.~A.\ 2013, \mnras, 436, 3430 
 
%\bibitem[Dale et al.(2014)]{2014arXiv1404.6102D} Dale, J.~E., Ngoumou, J., 
%Ercolano, B., \& Bonnell, I.~A.\ 2014, arXiv:1404.6102 

\bibitem[Dale et al.(2014)]{2014MNRAS.442..694D} Dale, J.~E., Ngoumou, J., 
Ercolano, B., \& Bonnell, I.~A.\ 2014, \mnras, 442, 694 

\bibitem[Deharveng et 
al.(2010)]{2010A&A...523A...6D} Deharveng, L., Schuller, F., Anderson, L.~D., et al.\ 2010, \aap, 523, A6

 \bibitem[Fierlinger et al.(2012)]{2012ASPC..453...25F} Fierlinger, K.~M., 
Burkert, A., Diehl, R., et al.\ 2012, Advances in Computational 
Astrophysics: Methods, Tools, and Outcome, 453, 25 

\bibitem[Freyer et al.(2003)]{2003ApJ...594..888F} Freyer, T., Hensler, G., 
\& Yorke, H.~W.\ 2003, \apj, 594, 888 
 
 \bibitem[Freyer et al.(2006)]{2006ApJ...638..262F} Freyer, T., Hensler, G., 
\& Yorke, H.~W.\ 2006, \apj, 638, 262 
 
\bibitem[Garcia-Segura 
\& Mac Low(1995)]{1995ApJ...455..145G} Garcia-Segura, G., \& Mac Low, M.-M.\ 1995, \apj, 455, 145 


\bibitem[Garcia-Segura et 
al.(1996a)]{1996A&A...316..133G} Garcia-Segura, G., Langer, N., \& Mac Low, M.-M.\ 1996b, \aap, 316, 133 

\bibitem[Garcia-Segura et 
al.(1996b)]{1996A&A...305..229G} Garcia-Segura, G., Mac Low, M.-M., \& Langer, N.\ 1996a, \aap, 305, 229 

\bibitem[G{\'o}rski et al.(2005)]{2005ApJ...622..759G} G{\'o}rski, K.~M., 
Hivon, E., Banday, A.~J., et al.\ 2005, \apj, 622, 759 

\bibitem[Gritschneder et al.(2010)]{2010ApJ...723..971G} Gritschneder, M., 
Burkert, A., Naab, T., \& Walch, S.\ 2010, \apj, 723, 971

 \bibitem[Gvaramadze et al.(2012)]{2012MNRAS.424.3037G} Gvaramadze, V.~V., 
Weidner, C., Kroupa, P., \& Pflamm-Altenburg, J.\ 2012, \mnras, 424, 3037

\bibitem[Harper-Clark 
\& Murray(2009)]{2009ApJ...693.1696H} Harper-Clark, E., \& Murray, N.\ 2009, \apj, 693, 1696

\bibitem[Hopkins et al.(2011)]{2011MNRAS.417..950H} Hopkins, P.~F., 
Quataert, E., \& Murray, N.\ 2011, \mnras, 417, 950 

\bibitem[Hubber et 
al.(2011)]{2011A&A...529A..27H} Hubber, D.~A., Batty, C.~P., McLeod, A., \& Whitworth, A.~P.\ 2011, \aap, 529, A27 

\bibitem[Hubber et al.(2013)]{2013MNRAS.430.3261H} Hubber, D.~A., Walch, 
S., \& Whitworth, A.~P.\ 2013, \mnras, 430, 3261 

%\bibitem[Kobulnicky 
%\& Fryer(2007)]{2007ApJ...670..747K} Kobulnicky, H.~A., \& Fryer, C.~L.\ 2007, \apj, 670, 747 

\bibitem[Kobulnicky et al.(2012)]{2012AJ....143...71K} Kobulnicky, H.~A., 
Lundquist, M.~J., Bhattacharjee, A., \& Kerton, C.~R.\ 2012, \aj, 143, 71 
 
%\bibitem[Krumholz 
%\& Thompson(2007)]{2007ApJ...669..289K} Krumholz, M.~R., \& Thompson, T.~A.\ 2007, \apj, 669, 289 

\bibitem[Krumholz et al.(2007)]{2007ApJ...671..518K} Krumholz, M.~R., 
Stone, J.~M., \& Gardiner, T.~A.\ 2007, \apj, 671, 518

%\bibitem[Lada 
%\& Lada(2003)]{2003ARA&A..41...57L} Lada, C.~J., \& Lada, E.~A.\ 2003, \araa, 41, 57 

\bibitem[Lamers 
\& Cassinelli(1999)]{1999isw..book.....L} Lamers, H.~J.~G.~L.~M., \& Cassinelli, J.~P.\ 1999, Introduction to Stellar Winds, by Henny J.~G.~L.~M.~Lamers and Joseph P.~Cassinelli, pp.~452.~ISBN 0521593980.~Cambridge, UK: Cambridge University Press, June 1999.,  

%\bibitem[Mac Low 
%\& Norman(1993)]{1993ApJ...407..207M} Mac Low, M.-M., \& Norman, M.~L.\ 1993, \apj, 407, 207 

\bibitem[Mac Low(2000)]{2000RMxAC...9..273M} Mac Low, M.-M.\ 2000, Revista 
Mexicana de Astronomia y Astrofisica Conference Series, 9, 273 

%\bibitem[Mackey et al.(2013)]{2013arXiv1308.5192M} Mackey, J., Langer, N., 
%\& Gvaramadze, V.~V.\ 2013, arXiv:1308.5192

\bibitem[Mackey et al.(2013)]{2013MNRAS.436..859M} Mackey, J., Langer, N., 
\& Gvaramadze, V.~V.\ 2013, \mnras, 436, 859

%\bibitem[Mason et al.(1998)]{1998AJ....115..821M} Mason, B.~D., Gies, 
%D.~R., Hartkopf, W.~I., et al.\ 1998, \aj, 115, 821 

\bibitem[Mohamed et 
al.(2012)]{2012A&A...541A...1M} Mohamed, S., Mackey, J., \& Langer, N.\ 2012, \aap, 541, A1 

\bibitem[Murray et al.(2011)]{2011ApJ...735...66M} Murray, N., M{\'e}nard, 
B., \& Thompson, T.~A.\ 2011, \apj, 735, 66 

\bibitem[Ngoumou et al.(2013)]{2013ApJ...769..139N} Ngoumou, J., Preibisch, 
T., Ratzka, T., \& Burkert, A.\ 2013, \apj, 769, 139

\bibitem[Ntormousi et al.(2011)]{2011ApJ...731...13N} Ntormousi, E., 
Burkert, A., Fierlinger, K., \& Heitsch, F.\ 2011, \apj, 731, 13 

\bibitem[Oey 
\& Garc{\'{\i}}a-Segura(2004)]{2004ApJ...613..302O} Oey, M.~S., \& Garc{\'{\i}}a-Segura, G.\ 2004, \apj, 613, 302 

\bibitem[Ohlendorf et 
al.(2012)]{2012A&A...540A..81O} Ohlendorf, H., Preibisch, T., Gaczkowski, B., et al.\ 2012, \aap, 540, A81

%\bibitem[Ohlendorf et 
%al.(2013)]{2013A&A...552A..14O} Ohlendorf, H., Preibisch, T., Gaczkowski, B., et al.\ 2013, \aap, 552, A14

\bibitem[Ostriker 
\& McKee(1988)]{1988RvMP...60....1O} Ostriker, J.~P., \& McKee, C.~F.\ 1988, Reviews of Modern Physics, 60, 1 

%\bibitem[Preibisch et al.(2001)]{2001IAUS..200...69P} Preibisch, T., 
%Weigelt, G., \& Zinnecker, H.\ 2001, The Formation of Binary Stars, 200, 69 

\bibitem[Preibisch et al.(2011)]{2011A&A...525A..92P} Preibisch, T., Schuller, F., Ohlendorf, H.,
et al.\ 2011a, \aap, 525, A92

\bibitem[Price(2007)]{2007PASA...24..159P} Price, D.~J.\ 2007, \pasa, 24, 
159 

\bibitem[Raga et al.(2012)]{2012RMxAA..48..199R} Raga, A.~C., Cant{\'o}, 
J., \& Rodr{\'{\i}}guez, L.~F.\ 2012, \rmxaa, 48, 199

\bibitem[Smith(2006)]{2006MNRAS.367..763S} Smith, N.\ 2006, \mnras, 367, 
763 

\bibitem[Spitzer(1978)]{1978ppim.book.....S} Spitzer, L.\ 1978, New York 
Wiley-Interscience, 1978.~333 p.,  

\bibitem[Steigman et al.(1975)]{1975ApJ...198..575S} Steigman, G., 
Strittmatter, P.~A., \& Williams, R.~E.\ 1975, \apj, 198, 575 

\bibitem[Str{\"o}mgren(1939)]{1939ApJ....89..526S} Str{\"o}mgren, B.\ 1939, 
\apj, 89, 526 

%\bibitem[Vishniac(1983)]{1983ApJ...274..152V} Vishniac, E.~T.\ 1983, \apj, 
%274, 152 

\bibitem[Walch et al.(2012)]{2012MNRAS.427..625W} Walch, S.~K., Whitworth, 
A.~P., Bisbas, T., W{\"u}nsch, R., \& Hubber, D.\ 2012, \mnras, 427, 625
  
 \bibitem[Weaver et al.(1977)]{1977ApJ...218..377W} Weaver, R., McCray, R., 
Castor, J., Shapiro, P., \& Moore, R.\ 1977, \apj, 218, 377 

  
\end{thebibliography}
\end{document}